%
\expandafter\ifx\csname phyzzx\endcsname\relax
 \message{It is better to use PHYZZX format than to
          \string\input\space PHYZZX}\else
 \wlog{PHYZZX macros are already loaded and are not
          \string\input\space again}%
 \endinput \fi
\catcode`\@=11 
\let\rel@x=\relax
\let\n@expand=\relax
\def\pr@tect{\let\n@expand=\noexpand}
\let\protect=\pr@tect
\let\gl@bal=\global
%
%
%
\newfam\cpfam
\newdimen\b@gheight             \b@gheight=12pt
\newcount\f@ntkey               \f@ntkey=0
\def\f@m{\afterassignment\samef@nt\f@ntkey=}
\def\samef@nt{\fam=\f@ntkey \the\textfont\f@ntkey\rel@x}
\def\setstr@t{\setbox\strutbox=\hbox{\vrule height 0.85\b@gheight
                                depth 0.35\b@gheight width\z@ }}
%
%
%
%
%

\font\fourteenrm  =cmr10 scaled\magstep2
\font\twelverm    =cmr12
\font\ninerm      =cmr9
\font\sixrm       =cmr6

\font\fourteenbf  =cmbx10 scaled\magstep2
\font\twelvebf    =cmbx12
\font\ninebf      =cmbx9
\font\sixbf       =cmbx6
\font\seventeeni  =cmmi10 scaled\magstep3    \skewchar\seventeeni='177
\font\fourteeni   =cmmi10 scaled\magstep2     \skewchar\fourteeni='177
\font\twelvei     =cmmi12                       \skewchar\twelvei='177
\font\ninei       =cmmi9                          \skewchar\ninei='177
\font\sixi        =cmmi6                           \skewchar\sixi='177
\font\seventeensy =cmsy10 scaled\magstep3    \skewchar\seventeensy='60
\font\fourteensy  =cmsy10 scaled\magstep2     \skewchar\fourteensy='60
\font\twelvesy    =cmsy10 scaled\magstep1       \skewchar\twelvesy='60
\font\ninesy      =cmsy9                          \skewchar\ninesy='60
\font\sixsy       =cmsy6                           \skewchar\sixsy='60

\font\fourteenex  =cmex10 scaled\magstep2
\font\twelveex    =cmex10 scaled\magstep1

\font\fourteensl  =cmsl10 scaled\magstep2
\font\twelvesl    =cmsl12
\font\ninesl      =cmsl9

\font\fourteenit  =cmti10 scaled\magstep2
\font\twelveit    =cmti12
\font\nineit      =cmti9
\font\fourteentt  =cmtt10 scaled\magstep2
\font\twelvett    =cmtt12
\font\fourteencp  =cmcsc10 scaled\magstep2
\font\twelvecp    =cmcsc10 scaled\magstep1
\font\tencp       =cmcsc10
%
%
\def\fourteenf@nts{\relax
    \textfont0=\fourteenrm          \scriptfont0=\tenrm
      \scriptscriptfont0=\sevenrm
    \textfont1=\fourteeni           \scriptfont1=\teni
      \scriptscriptfont1=\seveni
    \textfont2=\fourteensy          \scriptfont2=\tensy
      \scriptscriptfont2=\sevensy
    \textfont3=\fourteenex          \scriptfont3=\twelveex
      \scriptscriptfont3=\tenex
    \textfont\itfam=\fourteenit     \scriptfont\itfam=\tenit
    \textfont\slfam=\fourteensl     \scriptfont\slfam=\tensl
    \textfont\bffam=\fourteenbf     \scriptfont\bffam=\tenbf
      \scriptscriptfont\bffam=\sevenbf
    \textfont\ttfam=\fourteentt
    \textfont\cpfam=\fourteencp }
\def\twelvef@nts{\relax
    \textfont0=\twelverm          \scriptfont0=\ninerm
      \scriptscriptfont0=\sixrm
    \textfont1=\twelvei           \scriptfont1=\ninei
      \scriptscriptfont1=\sixi
    \textfont2=\twelvesy           \scriptfont2=\ninesy
      \scriptscriptfont2=\sixsy
    \textfont3=\twelveex          \scriptfont3=\tenex
      \scriptscriptfont3=\tenex
    \textfont\itfam=\twelveit     \scriptfont\itfam=\nineit
    \textfont\slfam=\twelvesl     \scriptfont\slfam=\ninesl
    \textfont\bffam=\twelvebf     \scriptfont\bffam=\ninebf
      \scriptscriptfont\bffam=\sixbf
    \textfont\ttfam=\twelvett
    \textfont\cpfam=\twelvecp }
\def\tenf@nts{\relax
    \textfont0=\tenrm          \scriptfont0=\sevenrm
      \scriptscriptfont0=\fiverm
    \textfont1=\teni           \scriptfont1=\seveni
      \scriptscriptfont1=\fivei
    \textfont2=\tensy          \scriptfont2=\sevensy
      \scriptscriptfont2=\fivesy
    \textfont3=\tenex          \scriptfont3=\tenex
      \scriptscriptfont3=\tenex
    \textfont\itfam=\tenit     \scriptfont\itfam=\seveni  
    \textfont\slfam=\tensl     \scriptfont\slfam=\sevenrm 
    \textfont\bffam=\tenbf     \scriptfont\bffam=\sevenbf
      \scriptscriptfont\bffam=\fivebf
    \textfont\ttfam=\tentt
    \textfont\cpfam=\tencp }
%
%

%
\def\rm{\n@expand\f@m0 }
\def\mit{\n@expand\f@m1 }         
\def\cal{\n@expand\f@m2 }
\def\it{\n@expand\f@m\itfam}
\def\sl{\n@expand\f@m\slfam}
\def\bf{\n@expand\f@m\bffam}
\def\tt{\n@expand\f@m\ttfam}
\def\caps{\n@expand\f@m\cpfam}    
\def\em@{\rel@x\ifnum\f@ntkey=0 \it \else
        \ifnum\f@ntkey=\bffam \it \else \rm \fi \fi }
\def\em{\n@expand\em@}
\def\fourteenpoint{\fourteenf@nts \samef@nt \b@gheight=14pt \setstr@t }
\def\twelvepoint{\twelvef@nts \samef@nt \b@gheight=12pt \setstr@t }
\def\tenpoint{\tenf@nts \samef@nt \b@gheight=10pt \setstr@t }
\normalbaselineskip = 19.2pt plus 0.2pt minus 0.1pt 
\normallineskip = 1.5pt plus 0.1pt minus 0.1pt
\normallineskiplimit = 1.5pt
\newskip\normaldisplayskip
\normaldisplayskip = 14.4pt plus 3.6pt minus 10.0pt 
\newskip\normaldispshortskip
\normaldispshortskip = 6pt plus 5pt
\newskip\normalparskip
\normalparskip = 6pt plus 2pt minus 1pt
\newskip\skipregister
\skipregister = 5pt plus 2pt minus 1.5pt
\newif\ifsingl@
\newif\ifdoubl@
\newif\iftwelv@  \twelv@true
\def\singlespace{\singl@true\doubl@false\spaces@t}
\def\doublespace{\singl@false\doubl@true\spaces@t}
\def\normalspace{\singl@false\doubl@false\spaces@t}
 \def\Tenpoint{\tenpoint\twelv@false\spaces@t}
\def\Twelvepoint{\twelvepoint\twelv@true\spaces@t}
\def\spaces@t{\rel@x
      \iftwelv@ \ifsingl@\subspaces@t3:4;\else\subspaces@t1:1;\fi
       \else \ifsingl@\subspaces@t3:5;\else\subspaces@t4:5;\fi \fi
      \ifdoubl@ \multiply\baselineskip by 5
         \divide\baselineskip by 4 \fi }
\def\subspaces@t#1:#2;{
      \baselineskip = \normalbaselineskip
      \multiply\baselineskip by #1 \divide\baselineskip by #2
      \lineskip = \normallineskip
      \multiply\lineskip by #1 \divide\lineskip by #2
      \lineskiplimit = \normallineskiplimit
      \multiply\lineskiplimit by #1 \divide\lineskiplimit by #2
      \parskip = \normalparskip
      \multiply\parskip by #1 \divide\parskip by #2
      \abovedisplayskip = \normaldisplayskip
      \multiply\abovedisplayskip by #1 \divide\abovedisplayskip by #2
      \belowdisplayskip = \abovedisplayskip
        \abovedisplayshortskip = \normaldispshortskip
      \multiply\abovedisplayshortskip by #1
        \divide\abovedisplayshortskip by #2
      \belowdisplayshortskip = \abovedisplayshortskip
      \advance\belowdisplayshortskip by \belowdisplayskip
      \divide\belowdisplayshortskip by 2
      \smallskipamount = \skipregister
      \multiply\smallskipamount by #1 \divide\smallskipamount by #2
      \medskipamount = \smallskipamount \multiply\medskipamount by 2
      \bigskipamount = \smallskipamount \multiply\bigskipamount by 4 }
\def\normalbaselines{ \baselineskip=\normalbaselineskip
   \lineskip=\normallineskip \lineskiplimit=\normallineskip
   \iftwelv@\else \multiply\baselineskip by 4 \divide\baselineskip by 5
     \multiply\lineskiplimit by 4 \divide\lineskiplimit by 5
     \multiply\lineskip by 4 \divide\lineskip by 5 \fi }
\Twelvepoint  
\interlinepenalty=50
\interfootnotelinepenalty=5000
 \predisplaypenalty=9000
\postdisplaypenalty=500
\hfuzz=1pt
\vfuzz=0.2pt
\newdimen\HOFFSET  \HOFFSET=0pt
\newdimen\VOFFSET  \VOFFSET=0pt
\newdimen\HSWING   \HSWING=0pt
\dimen\footins=8in
%
%
%
\newskip\pagebottomfiller
\pagebottomfiller=\z@ plus \z@ minus \z@
\def\pagecontents{
   \ifvoid\topins\else\unvbox\topins\vskip\skip\topins\fi
   \dimen@ = \dp255 \unvbox255
   \vskip\pagebottomfiller
   \ifvoid\footins\else\vskip\skip\footins\footrule\unvbox\footins\fi
    \ifr@ggedbottom \kern-\dimen@ \vfil \fi }
\def\makeheadline{\vbox to 0pt{ \skip@=\topskip
      \advance\skip@ by -12pt \advance\skip@ by -2\normalbaselineskip
      \vskip\skip@ \line{\vbox to 12pt{}\the\headline} \vss
      }\nointerlineskip}
\def\makefootline{\baselineskip = 1.5\normalbaselineskip
                 \line{\the\footline}}
\newif\iffrontpage
\newif\ifp@genum
\def\nopagenumbers{\p@genumfalse}
\def\pagenumbers{\p@genumtrue}
\pagenumbers
\newtoks\paperheadline
\newtoks\paperfootline
\newtoks\letterheadline
\newtoks\letterfootline
\newtoks\letterinfo
\newtoks\date
\paperheadline={\hfil}
\paperfootline={\hss\iffrontpage\else\ifp@genum\tenrm\folio\hss\fi\fi}
\letterheadline{\iffrontpage \hfil \else
    \rm \ifp@genum page~~\folio\fi \hfil\the\date \fi}
\letterfootline={\iffrontpage\the\letterinfo\else\hfil\fi}
\letterinfo={\hfil}
\def\monthname{\rel@x\ifcase\month 0/\or January\or February\or
   March\or April\or May\or June\or July\or August\or September\or
   October\or November\or December\else\number\month/\fi}
\def\today{\monthname~\number\day, \number\year}
\date={\today}
\headline=\paperheadline 
\footline=\paperfootline 
\countdef\pageno=1      \countdef\pagen@=0
\countdef\pagenumber=1  \pagenumber=1
\def\advancepageno{\gl@bal\advance\pagen@ by 1
   \ifnum\pagenumber<0 \gl@bal\advance\pagenumber by -1
     \else\gl@bal\advance\pagenumber by 1 \fi
    \gl@bal\frontpagefalse  \swing@ }
\def\folio{\ifnum\pagenumber<0 \romannumeral-\pagenumber
           \else \number\pagenumber \fi }
\def\swing@{\ifodd\pagenumber \gl@bal\advance\hoffset by -\HSWING
             \else \gl@bal\advance\hoffset by \HSWING \fi }
\def\footrule{\dimen@=\prevdepth\nointerlineskip
   \vbox to 0pt{\vskip -0.25\baselineskip \hrule width 0.35\hsize \vss}
   \prevdepth=\dimen@ }
\let\footnotespecial=\rel@x
\newdimen\footindent
\footindent=24pt
\def\Textindent#1{\noindent\llap{#1\enspace}\ignorespaces}
\def\Vfootnote#1{\insert\footins\bgroup
   \interlinepenalty=\interfootnotelinepenalty \floatingpenalty=20000
   \singl@true\doubl@false\Tenpoint
  \splittopskip=\ht\strutbox \boxmaxdepth=\dp\strutbox
   \leftskip=\footindent \rightskip=\z@skip
   \parindent=0.5\footindent \parfillskip=0pt plus 1fil
   \spaceskip=\z@skip \xspaceskip=\z@skip \footnotespecial
   \Textindent{#1}\footstrut\futurelet\next\fo@t}

\def\vfootnote#1{\Vfootnote{${#1}$}}
\def\footnote#1{\attach{#1}\vfootnote{#1}}

\let\footsymbol=\star
\newcount\lastf@@t           \lastf@@t=-1
\newcount\footsymbolcount    \footsymbolcount=0
\newif\ifPhysRev
\def\bumpfootsymbolcount{\rel@x
   \iffrontpage \bumpfootsymbolpos \else \advance\lastf@@t by 1
  \iffrontpage \bumpfootsymbolpos \else \advance\lastf@@t by 1
     \ifPhysRev \bumpfootsymbolneg \else \bumpfootsymbolpos \fi \fi
   \gl@bal\lastf@@t=\pagen@ }
\def\bumpfootsymbolpos{\ifnum\footsymbolcount <0
                            \gl@bal\footsymbolcount =0 \fi
    \ifnum\lastf@@t<\pagen@ \gl@bal\footsymbolcount=0
     \else \gl@bal\advance\footsymbolcount by 1 \fi }
\def\bumpfootsymbolneg{\ifnum\footsymbolcount >0
             \gl@bal\footsymbolcount =0 \fi
         \gl@bal\advance\footsymbolcount by -1 }
\def\fd@f#1 {\xdef\footsymbol{\mathchar"#1 }}
\def\generatefootsymbol{\ifcase\footsymbolcount \fd@f 13F \or \fd@f 279
        \or \fd@f 27A \or \fd@f 278 \or \fd@f 27B \else
        \ifnum\footsymbolcount <0 \fd@f{023 \number-\footsymbolcount }
         \else \fd@f 203 {\loop \ifnum\footsymbolcount >5
                \fd@f{203 \footsymbol } \advance\footsymbolcount by -1
                \repeat }\fi \fi }

\def\nonfrenchspacing{\sfcode`\.=3001 \sfcode`\!=3000 \sfcode`\?=3000
        \sfcode`\:=2000 \sfcode`\;=1500 \sfcode`\,=1251 }
\nonfrenchspacing
\newdimen\d@twidth
{\setbox0=\hbox{s.} \gl@bal\d@twidth=\wd0 \setbox0=\hbox{s}
        \gl@bal\advance\d@twidth by -\wd0 }
\def\removehglue{\loop \unskip \ifdim\lastskip >\z@ \repeat }
\def\roll@ver#1{\removehglue \nobreak \count255 =\spacefactor \dimen@=\z@
        \ifnum\count255 =3001 \dimen@=\d@twidth \fi
        \ifnum\count255 =1251 \dimen@=\d@twidth \fi
    \iftwelv@ \kern-\dimen@ \else \kern-0.83\dimen@ \fi
   #1\spacefactor=\count255 }
\def\step@ver#1{\rel@x \ifmmode #1\else \ifhmode
        \roll@ver{${}#1$}\else {\setbox0=\hbox{${}#1$}}\fi\fi }
\def\attach#1{\step@ver{\strut^{\mkern 2mu #1} }}
%
%
%
\newcount\chapternumber      \chapternumber=0
\newcount\sectionnumber      \sectionnumber=0
\newcount\equanumber         \equanumber=0
\let\chapterlabel=\rel@x
\let\sectionlabel=\rel@x
\newtoks\chapterstyle        \chapterstyle={\Number}
\newtoks\sectionstyle        \sectionstyle={\chapterlabel.\Number}
\newskip\chapterskip         \chapterskip=\bigskipamount
\newskip\sectionskip         \sectionskip=\medskipamount
\newskip\headskip            \headskip=8pt plus 3pt minus 3pt
\newdimen\chapterminspace    \chapterminspace=15pc
\newdimen\sectionminspace    \sectionminspace=10pc
\newdimen\referenceminspace  \referenceminspace=20pc
\def\chapterreset{\gl@bal\advance\chapternumber by 1
   \ifnum\equanumber<0 \else\gl@bal\equanumber=0\fi
   \sectionnumber=0 \let\sectionlabel=\rel@x
   {\pr@tect\xdef\chapterlabel{\the\chapterstyle{\the\chapternumber}}}}
\def\alphabetic#1{\count255='140 \advance\count255 by #1\char\count255}
\def\Alphabetic#1{\count255='100 \advance\count255 by #1\char\count255}
\def\Roman#1{\uppercase\expandafter{\romannumeral #1}}
\def\roman#1{\romannumeral #1}
\def\Number#1{\number #1}
\def\BLANC#1{}
\def\titleparagraphs{\interlinepenalty=9999
     \leftskip=0.03\hsize plus 0.22\hsize minus 0.03\hsize
     \rightskip=\leftskip \parfillskip=0pt
     \hyphenpenalty=9000 \exhyphenpenalty=9000
     \tolerance=9999 \pretolerance=9000
     \spaceskip=0.333em \xspaceskip=0.5em }
\def\titlestyle#1{\par\begingroup \titleparagraphs
     \iftwelv@\fourteenpoint\else\twelvepoint\fi
   \noindent #1\par\endgroup }
\def\spacecheck#1{\dimen@=\pagegoal\advance\dimen@ by -\pagetotal
   \ifdim\dimen@<#1 \ifdim\dimen@>0pt \vfil\break \fi\fi}
\def\chapter#1{\par \penalty-300 \vskip\chapterskip
   \spacecheck\chapterminspace
   \chapterreset \titlestyle{\chapterlabel.~#1}
   \nobreak\vskip\headskip \penalty 30000
   {\pr@tect\wlog{\string\chapter\space \chapterlabel}} }

\def\section#1{\par \ifnum\the\lastpenalty=30000\else
   \penalty-200\vskip\sectionskip \spacecheck\sectionminspace\fi
   \gl@bal\advance\sectionnumber by 1
   {\pr@tect
   \xdef\sectionlabel{\the\sectionstyle\the\sectionnumber}
   \wlog{\string\section\space \sectionlabel}}
   \noindent {\caps\enspace\sectionlabel.~~#1}\par
   \nobreak\vskip\headskip \penalty 30000 }
\def\subsection#1{\par
   \ifnum\the\lastpenalty=30000\else \penalty-100\smallskip \fi
   \noindent\undertext{#1}\enspace \vadjust{\penalty5000}}

\def\undertext#1{\vtop{\hbox{#1}\kern 1pt \hrule}}
\def\ACK{\par\penalty-100\medskip \spacecheck\sectionminspace
   \line{\fourteenrm\hfil ACKNOWLEDGEMENTS\hfil}\nobreak\vskip\headskip }

\def\APPENDIX#1#2{\par\penalty-300\vskip\chapterskip
   \spacecheck\chapterminspace \chapterreset \xdef\chapterlabel{#1}
   \titlestyle{APPENDIX #2} \nobreak\vskip\headskip \penalty 30000
   \wlog{\string\Appendix~\chapterlabel} }
\def\Appendix#1{\APPENDIX{#1}{#1}}
\def\appendix{\APPENDIX{A}{}}
\def\unnumberedchapters{\let\makechapterlabel=\rel@x
      \let\chapterlabel=\rel@x  \sectionstyle={\BLANC}
      \let\sectionlabel=\rel@x \sequentialequations }
%
%
%
\def\eqname#1{\rel@x {\pr@tect
  \ifnum\equanumber<0 \xdef#1{{\rm(\number-\equanumber)}}%
     \gl@bal\advance\equanumber by -1
  \else \gl@bal\advance\equanumber by 1
     \ifx\chapterlabel\rel@x \def\d@t{}\else \def\d@t{.}\fi
    \xdef#1{{\rm(\chapterlabel\d@t\number\equanumber)}}\fi #1}}
\def\eqinsert#1{\noalign{\dimen@=\prevdepth \nointerlineskip
   \setbox0=\hbox to\displaywidth{\hfil #1}
   \vbox to 0pt{\kern 0.5\baselineskip\hbox{$\!\box0\!$}\vss}
   \prevdepth=\dimen@}}
%

%
%
\def\GENITEM#1;#2{\par \hangafter=0 \hangindent=#1
    \Textindent{$ #2 $}\ignorespaces}
\outer\def\newitem#1=#2;{\gdef#1{\GENITEM #2;}}

\newdimen\itemsize                \itemsize=30pt
\newitem\item=1\itemsize;
\newitem\sitem=1.75\itemsize;     
\newitem\ssitem=2.5\itemsize;     
\outer\def\newlist#1=#2&#3&#4;{\toks0={#2}\toks1={#3}%
   \count255=\escapechar \escapechar=-1
   \alloc@0\list\countdef\insc@unt\listcount     \listcount=0
   \edef#1{\par
      \countdef\listcount=\the\allocationnumber
      \advance\listcount by 1
      \hangafter=0 \hangindent=#4
      \Textindent{\the\toks0{\listcount}\the\toks1}}
   \expandafter\expandafter\expandafter
    \edef\c@t#1{begin}{\par
      \countdef\listcount=\the\allocationnumber \listcount=1
      \hangafter=0 \hangindent=#4
      \Textindent{\the\toks0{\listcount}\the\toks1}}
   \expandafter\expandafter\expandafter
    \edef\c@t#1{con}{\par \hangafter=0 \hangindent=#4 \noindent}
   \escapechar=\count255}
  \def\c@t#1#2{\csname\string#1#2\endcsname}
\newlist\point=\Number&.&1.0\itemsize;
\newlist\subpoint=(\alphabetic&)&1.75\itemsize;
\newlist\subsubpoint=(\roman&)&2.5\itemsize;
%

%
%
%
%
\newcount\referencecount     \referencecount=0
\newcount\lastrefsbegincount \lastrefsbegincount=0
\newif\ifreferenceopen       \newwrite\referencewrite
\newdimen\refindent          \refindent=30pt
\def\normalrefmark#1{\attach{\scriptscriptstyle [ #1 ] }}
\let\PRrefmark=\attach
\def\NPrefmark#1{\step@ver{{\;[#1]}}}
\def\refmark#1{\rel@x\ifPhysRev\PRrefmark{#1}\else\normalrefmark{#1}\fi}
\def\refend@{\refmark{\number\referencecount}}
\def\refend{\refend@{}\space }
\def\refsend{\refmark{\count255=\referencecount
   \advance\count255 by-\lastrefsbegincount
   \ifcase\count255 \number\referencecount
   \or \number\lastrefsbegincount,\number\referencecount
   \else \number\lastrefsbegincount-\number\referencecount \fi}\space }
\def\REFNUM#1{\rel@x \gl@bal\advance\referencecount by 1
    \xdef#1{\the\referencecount }}
\def\Refnum#1{\REFNUM #1\refend@ } 
\def\REF#1{\REFNUM #1\R@FWRITE\ignorespaces}
\def\Ref#1{\Refnum #1\REFWRITE }
\def\ref{\Ref\?}
\def\REFS#1{\REFNUM #1\gl@bal\lastrefsbegincount=\referencecount
    \REFWRITE }

\def\r@fitem#1{\par \hangafter=0 \hangindent=\refindent \Textindent{#1}}
\def\refitem#1{\r@fitem{#1.}}
\def\NPrefitem#1{\r@fitem{[#1]}}
\def\NPrefs{\let\refmark=\NPrefmark \let\refitem=\NPrefitem}
\def\REFWRITE{\R@FWRITE\rel@x }
\def\R@FWRITE#1{\ifreferenceopen \else \gl@bal\referenceopentrue
     \immediate\openout\referencewrite=\jobname.refs
     \toks@={\begingroup \refoutspecials \catcode`\^^M=10 }%
     \immediate\write\referencewrite{\the\toks@}\fi
    \immediate\write\referencewrite{\noexpand\refitem %
                                    {\the\referencecount}}%
    \p@rse@ndwrite \referencewrite #1}
\begingroup
 \catcode`\^^M=\active \let^^M=\relax %
 \gdef\p@rse@ndwrite#1#2{\begingroup \catcode`\^^M=12 \newlinechar=`\^^M%
         \chardef\rw@write=#1\sc@nlines#2}%
 \gdef\sc@nlines#1#2{\sc@n@line \g@rbage #2^^M\endsc@n \endgroup #1}%
 \gdef\sc@n@line#1^^M{\expandafter\toks@\expandafter{\deg@rbage #1}%
         \immediate\write\rw@write{\the\toks@}%
         \futurelet\n@xt \sc@ntest }%
\endgroup
\def\sc@ntest{\ifx\n@xt\endsc@n \let\n@xt=\rel@x
       \else \let\n@xt=\sc@n@notherline \fi \n@xt }
\def\sc@n@notherline{\sc@n@line \g@rbage }
\def\deg@rbage#1{}
\let\g@rbage=\relax    \let\endsc@n=\relax
\def\refout{\par\penalty-400\vskip\chapterskip
   \spacecheck\referenceminspace
   \ifreferenceopen \Closeout\referencewrite \referenceopenfalse \fi
   \line{\fourteenrm\hfil REFERENCES\hfil}\vskip\headskip
   \input \jobname.refs
   }
\def\refoutspecials{\sfcode`\.=1000 \interlinepenalty=1000
         \rightskip=\z@ plus 1em minus \z@ }
\def\Closeout#1{\toks0={\par\endgroup}\immediate\write#1{\the\toks0}%
   \immediate\closeout#1}
%
%
\newcount\figurecount     \figurecount=0
\newcount\tablecount      \tablecount=0
\newif\iffigureopen       \newwrite\figurewrite
\newif\iftableopen        \newwrite\tablewrite
\def\FIGNUM#1{\rel@x \gl@bal\advance\figurecount by 1
    \xdef#1{\the\figurecount}}
\def\FIGURE#1{\FIGNUM #1\F@GWRITE\ignorespaces }

\def\figitem#1{\r@fitem{#1)}}
\def\FIGWRITE{\F@GWRITE\rel@x }
\def\TABNUM#1{\rel@x \gl@bal\advance\tablecount by 1
    \xdef#1{\the\tablecount}}
\def\TABLE#1{\TABNUM #1\T@BWRITE\ignorespaces }

\def\tabitem#1{\r@fitem{#1:}}
\def\TABWRITE{\T@BWRITE\rel@x }
\def\F@GWRITE#1{\iffigureopen \else \gl@bal\figureopentrue
     \immediate\openout\figurewrite=\jobname.figs
     \toks@={\begingroup \catcode`\^^M=10 }%
     \immediate\write\figurewrite{\the\toks@}\fi
    \immediate\write\figurewrite{\noexpand\figitem %
                                 {\the\figurecount}}%
    \p@rse@ndwrite \figurewrite #1}
\def\T@BWRITE#1{\iftableopen \else \gl@bal\tableopentrue
     \immediate\openout\tablewrite=\jobname.tabs
     \toks@={\begingroup \catcode`\^^M=10 }%
     \immediate\write\tablewrite{\the\toks@}\fi
    \immediate\write\tablewrite{\noexpand\tabitem %
                                 {\the\tablecount}}%
    \p@rse@ndwrite \tablewrite #1}
\def\figout{\par\penalty-400
   \vskip\chapterskip\spacecheck\referenceminspace
   \iffigureopen \Closeout\figurewrite \figureopenfalse \fi
   \line{\fourteenrm\hfil FIGURE CAPTIONS\hfil}\vskip\headskip
   \input \jobname.figs
   }
\def\tabout{\par\penalty-400
   \vskip\chapterskip\spacecheck\referenceminspace
   \iftableopen \Closeout\tablewrite \tableopenfalse \fi
   \line{\fourteenrm\hfil TABLE CAPTIONS\hfil}\vskip\headskip
   \input \jobname.tabs
   }
%
%
%
\newbox\picturebox
\def\p@cht{\ht\picturebox }
\def\p@cwd{\wd\picturebox }
\def\p@cdp{\dp\picturebox }
\newdimen\xshift
\newdimen\yshift
\newdimen\captionwidth
\newskip\captionskip
\captionskip=15pt plus 5pt minus 3pt
\def\fullwidth{\captionwidth=\hsize }
\newtoks\Caption
\newif\ifcaptioned
\newif\ifselfcaptioned
\def\caption{\captionedtrue \Caption }
\newcount\linesabove
\newif\iffileexists
\newtoks\picfilename
\def\fil@#1 {\fileexiststrue \picfilename={#1}}
\def\file#1{\if=#1\let\n@xt=\fil@ \else \def\n@xt{\fil@ #1}\fi \n@xt }
\def\pl@t{\begingroup \pr@tect
    \setbox\picturebox=\hbox{}\fileexistsfalse
    \let\height=\p@cht \let\width=\p@cwd \let\depth=\p@cdp
    \xshift=\z@ \yshift=\z@ \captionwidth=\z@
    \Caption={}\captionedfalse
    \linesabove =0 \picturedefault }
\def\plot{\pl@t \selfcaptionedfalse }
\def\Picture#1{\gl@bal\advance\figurecount by 1
    \xdef#1{\the\figurecount}\pl@t \selfcaptionedtrue }

\def\s@vepicture{\iffileexists \parsefilename \redopicturebox \fi
   \ifdim\captionwidth>\z@ \else \captionwidth=\p@cwd \fi
   \xdef\lastpicture{\iffileexists
        \setbox0=\hbox{\raise\the\yshift \vbox{%
              \moveright\the\xshift\hbox{\picturedefinition}}}%
        \else \setbox0=\hbox{}\fi
         \ht0=\the\p@cht \wd0=\the\p@cwd \dp0=\the\p@cdp
         \vbox{\hsize=\the\captionwidth \line{\hss\box0 \hss }%
              \ifcaptioned \vskip\the\captionskip \noexpand\Tenpoint
                \ifselfcaptioned Figure~\the\figurecount.\enspace \fi
                \the\Caption \fi }}%
    \endgroup }
\let\endpicture=\s@vepicture
\def\savepicture#1{\s@vepicture \global\let#1=\lastpicture }
\def\displaypicture{\fullwidth \s@vepicture $$\lastpicture $${}}
\def\toppicture{\fullwidth \s@vepicture \topinsert
    \lastpicture \medskip \endinsert }
\def\midpicture{\fullwidth \s@vepicture \midinsert
    \lastpicture \endinsert }
%
%
\def\leftpicture{\pres@tpicture
    \dimen@i=\hsize \advance\dimen@i by -\dimen@ii
    \setbox\picturebox=\hbox to \hsize {\box0 \hss }%
    \wr@paround }
\def\rightpicture{\pres@tpicture
    \dimen@i=\z@
    \setbox\picturebox=\hbox to \hsize {\hss \box0 }%
    \wr@paround }
\def\pres@tpicture{\gl@bal\linesabove=\linesabove
    \s@vepicture \setbox\picturebox=\vbox{
         \kern \linesabove\baselineskip \kern 0.3\baselineskip
         \lastpicture \kern 0.3\baselineskip }%
    \dimen@=\p@cht \dimen@i=\dimen@
    \advance\dimen@i by \pagetotal
    \par \ifdim\dimen@i>\pagegoal \vfil\break \fi
    \dimen@ii=\hsize
    \advance\dimen@ii by -\parindent \advance\dimen@ii by -\p@cwd
    \setbox0=\vbox to\z@{\kern-\baselineskip \unvbox\picturebox \vss }}
\def\wr@paround{\Caption={}\count255=1
    \loop \ifnum \linesabove >0
         \advance\linesabove by -1 \advance\count255 by 1
         \advance\dimen@ by -\baselineskip
         \expandafter\Caption \expandafter{\the\Caption \z@ \hsize }%
      \repeat
    \loop \ifdim \dimen@ >\z@
         \advance\count255 by 1 \advance\dimen@ by -\baselineskip
         \expandafter\Caption \expandafter{%
             \the\Caption \dimen@i \dimen@ii }%
      \repeat
    \edef\n@xt{\parshape=\the\count255 \the\Caption \z@ \hsize }%
    \par\noindent \n@xt \strut \vadjust{\box\picturebox }}
\let\picturedefault=\relax
\let\parsefilename=\relax
\def\redopicturebox{\let\picturedefinition=\rel@x
   \errhelp=\disabledpictures
   \errmessage{This version of TeX cannot handle pictures.  Sorry.}}
\newhelp\disabledpictures
     {You will get a blank box in place of your picture.}
%
%
%
%
%
%
%
%
%
%
\def\FRONTPAGE{\ifvoid255\else\vfill\penalty-20000\fi
   \gl@bal\pagenumber=1     \gl@bal\chapternumber=0
   \gl@bal\equanumber=0     \gl@bal\sectionnumber=0
   \gl@bal\referencecount=0 \gl@bal\figurecount=0
   \gl@bal\tablecount=0     \gl@bal\frontpagetrue
   \gl@bal\lastf@@t=0       \gl@bal\footsymbolcount=0}

\def\papers{\papersize\headline=\paperheadline\footline=\paperfootline}
\def\papersize{
   \advance\hoffset by\HOFFSET \advance\voffset by\VOFFSET
   \pagebottomfiller=0pc
   \skip\footins=\bigskipamount \normalspace }
\papers  
%
%
\newskip\lettertopskip       \lettertopskip=20pt plus 50pt
\newskip\letterbottomskip    \letterbottomskip=\z@ plus 100pt
\newskip\signatureskip       \signatureskip=40pt plus 3pt
\def\lettersize{\hsize=6.5in \vsize=8.5in \hoffset=0in \voffset=0.5in
   \advance\hoffset by\HOFFSET \advance\voffset by\VOFFSET
   \pagebottomfiller=\letterbottomskip
   \skip\footins=\smallskipamount \multiply\skip\footins by 3
   \singlespace }
\def\MEMO{\lettersize \headline=\letterheadline \footline={\hfil }%
   \let\rule=\memorule \FRONTPAGE \memohead }

\def\memodate{\afterassignment\MEMO \date }
\def\memit@m#1{\smallskip \hangafter=0 \hangindent=1in
    \Textindent{\caps #1}}
\def\subject{\memit@m{Subject:}}
\def\topic{\memit@m{Topic:}}
\def\from{\memit@m{From:}}
\def\memorule{\medskip\hrule height 1pt\bigskip}  
\def\memohead{\centerline{\fourteenrm MEMORANDUM}}
\newwrite\labelswrite
\newtoks\rw@toks
\def\letters{\lettersize
   \headline=\letterheadline \footline=\letterfootline
   \immediate\openout\labelswrite=\jobname.lab}

\let\letterhead=\rel@x
\def\addressee#1{\medskip\line{\hskip 0.75\hsize plus\z@ minus 0.25\hsize
                               \the\date \hfil }%
   \vskip \lettertopskip
   \ialign to\hsize{\strut ##\hfil\tabskip 0pt plus \hsize \crcr #1\crcr}
   \writelabel{#1}\medskip \noindent\hskip -\spaceskip \ignorespaces }
\def\rwl@begin#1\cr{\rw@toks={#1\crcr}\rel@x
   \immediate\write\labelswrite{\the\rw@toks}\futurelet\n@xt\rwl@next}
\def\rwl@next{\ifx\n@xt\rwl@end \let\n@xt=\rel@x
      \else \let\n@xt=\rwl@begin \fi \n@xt}
\let\rwl@end=\rel@x
\def\writelabel#1{\immediate\write\labelswrite{\noexpand\labelbegin}
     \rwl@begin #1\cr\rwl@end
     \immediate\write\labelswrite{\noexpand\labelend}}
\newtoks\FromAddress         \FromAddress={}
\newtoks\sendername          \sendername={}
\newbox\FromLabelBox
\newdimen\labelwidth          \labelwidth=6in
\def\makelabels{\afterassignment\Makelabels \sendersname=}
\def\Makelabels{\FRONTPAGE \letterinfo={\hfil } \MakeFromBox
     \immediate\closeout\labelswrite  \input \jobname.lab\vfil\eject}
\let\labelend=\rel@x
\def\labelbegin#1\labelend{\setbox0=\vbox{\ialign{##\hfil\cr #1\crcr}}
     \MakeALabel }
\def\MakeFromBox{\gl@bal\setbox\FromLabelBox=\vbox{\Tenpoint
     \ialign{##\hfil\cr \the\sendername \the\FromAddress \crcr }}}
\def\MakeALabel{\vskip 1pt \hbox{\vrule \vbox{
        \hsize=\labelwidth \hrule\bigskip
        \leftline{\hskip 1\parindent \copy\FromLabelBox}\bigskip
        \centerline{\hfil \box0 } \bigskip \hrule
        }\vrule } \vskip 1pt plus 1fil }
\def\signed#1{\par \nobreak \bigskip \dt@pfalse \begingroup
  \everycr={\noalign{\nobreak
            \ifdt@p\vskip\signatureskip\gl@bal\dt@pfalse\fi }}%
  \tabskip=0.5\hsize plus \z@ minus 0.5\hsize
  \halign to\hsize {\strut ##\hfil\tabskip=\z@ plus 1fil minus \z@\crcr
          \noalign{\gl@bal\dt@ptrue}#1\crcr }%
  \endgroup \bigskip }
\newbox\letterb@x
\def\lettertext{\par \vskip\parskip \unvcopy\letterb@x \par }
\def\multiletter{\setbox\letterb@x=\vbox\bgroup
      \everypar{\vrule height 1\baselineskip depth 0pt width 0pt }
      \singlespace \topskip=\baselineskip }
\def\letterend{\par\egroup}
%
%
%
\newskip\frontpageskip
\newtoks\Pubnum   
\newtoks\Pubtype  \let\pubtype=\Pubtype
\newif\ifp@bblock  \p@bblocktrue
\def\PH@SR@V{\doubl@true \baselineskip=24.1pt plus 0.2pt minus 0.1pt
             \parskip= 3pt plus 2pt minus 1pt }
\def\PHYSREV{\papers\PhysRevtrue\PH@SR@V}

\def\titlepage{\FRONTPAGE\papers\ifPhysRev\PH@SR@V\fi
   \ifp@bblock\p@bblock \else\hrule height\z@ \rel@x \fi }
\def\nopubblock{\p@bblockfalse}
\def\endpage{\vfil\break}
\frontpageskip=12pt plus .5fil minus 2pt
\Pubtype={}
\Pubnum={}
\def\p@bblock{\begingroup \tabskip=\hsize minus \hsize
   \baselineskip=1.5\ht\strutbox \topspace-2\baselineskip
   \halign to\hsize{\strut ##\hfil\tabskip=0pt\crcr
       \the\Pubnum\crcr\the\date\crcr\the\pubtype\crcr}\endgroup}
\def\title#1{\vskip\frontpageskip \titlestyle{#1} \vskip\headskip }
\def\author#1{\vskip\frontpageskip\titlestyle{\twelvecp #1}\nobreak}

\def\address#1{\par\kern 5pt\titlestyle{\twelvepoint\it #1}}
\def\andaddress{\par\kern 5pt \centerline{\sl and} \address}

\def\abstract{\par\dimen@=\prevdepth \hrule height\z@ \prevdepth=\dimen@
   \vskip\frontpageskip\centerline{\fourteenrm ABSTRACT}\vskip\headskip }

%
%
%

\def\\{\rel@x \ifmmode \backslash \else {\tt\char`\\}\fi }
\def\sequentialequations{\rel@x \if\equanumber<0 \else
  \gl@bal\equanumber=-\equanumber \gl@bal\advance\equanumber by -1 \fi }
\def\journal#1&#2(#3){\begingroup \let\journal=\dummyj@urnal
    \unskip, \sl #1\unskip~\bf\ignorespaces #2\rm
    (\afterassignment\j@ur \count255=#3), \endgroup\ignorespaces }
\def\j@ur{\ifnum\count255<100 \advance\count255 by 1900 \fi
          \number\count255 }
\def\dummyj@urnal{%
    \toks@={Reference foul up: nested \journal macros}%
    \errhelp={Your forgot & or ( ) after the last \journal}%
    \errmessage{\the\toks@ }}

\def\topspace{\hrule height 0pt depth 0pt \vskip}

\def\Buildrel#1\under#2{\mathrel{\mathop{#2}\limits_{#1}}}
\def\becomes#1{\mathchoice{\becomes@\scriptstyle{#1}}
   {\becomes@\scriptstyle{#1}} {\becomes@\scriptscriptstyle{#1}}
   {\becomes@\scriptscriptstyle{#1}}}
\def\becomes@#1#2{\mathrel{\setbox0=\hbox{$\m@th #1{\,#2\,}$}%
        \mathop{\hbox to \wd0 {\rightarrowfill}}\limits_{#2}}}

\let\int=\intop         
\def\lsim{\mathrel{\mathpalette\@versim<}}
\def\gsim{\mathrel{\mathpalette\@versim>}}
\def\@versim#1#2{\vcenter{\offinterlineskip
        \ialign{$\m@th#1\hfil##\hfil$\crcr#2\crcr\sim\crcr } }}
\def\big#1{{\hbox{$\left#1\vbox to 0.85\b@gheight{}\right.\n@space$}}}
\def\Big#1{{\hbox{$\left#1\vbox to 1.15\b@gheight{}\right.\n@space$}}}
\def\bigg#1{{\hbox{$\left#1\vbox to 1.45\b@gheight{}\right.\n@space$}}}
\def\Bigg#1{{\hbox{$\left#1\vbox to 1.75\b@gheight{}\right.\n@space$}}}
\def\){\mskip 2mu\nobreak }
%
%
%
\let\sec@nt=\sec
\def\sec{\rel@x\ifmmode\let\n@xt=\sec@nt\else\let\n@xt\section\fi\n@xt}
\def\obsolete#1{\message{Macro \string #1 is obsolete.}}
\def\firstsec#1{\obsolete\firstsec \section{#1}}
\def\firstsubsec#1{\obsolete\firstsubsec \subsection{#1}}
\def\thispage#1{\obsolete\thispage \gl@bal\pagenumber=#1\frontpagefalse}
\def\thischapter#1{\obsolete\thischapter \gl@bal\chapternumber=#1}
\def\splitout{\obsolete\splitout\rel@x}
 \def\prop{\obsolete\prop \propto }
\def\nextequation#1{\obsolete\nextequation \gl@bal\equanumber=#1
   \ifnum\the\equanumber>0 \gl@bal\advance\equanumber by 1 \fi}
\def\BOXITEM{\afterassigment\B@XITEM\setbox0=}
\def\B@XITEM{\par\hangindent\wd0 \noindent\box0 }
%
%
%
\def\phyzzx{PHY\setbox0=\hbox{Z}\copy0 \kern-0.5\wd0 \box0 X}
        
\everyjob{\xdef\today{\monthname~\number\day, \number\year}
        \input myphyx.tex }
\message{ by V.K.}
%
\catcode`\@=12 
%

\singlespace
\overfullrule 0pt
\baselineskip=18pt
\gdef\journal #1, #2, #3, 1#4#5#6{
      {\sl #1~}{\bf #2}, #3 (1#4#5#6)}

\def\PRB{\journal Phys. Rev. B, }

\def\NPB{\journal Nucl. Phys. B., }

\def\PRL{\journal Phys. Rev. Lett., }

\def\Euphys{\journal Europhys. Lett., }
\def\ltwid{\mathrel{\raise.3ex\hbox{$<$\kern-.75em\lower1ex\hbox{$\sim$}}}}
\def\bfx{{\bf x}}
\def\bfy{{\bf y}}

\def\bfq{{\bf q}}

\def\sh{\rm sh}
%
\titlepage
\title{{\bf Coulomb drag in double layer electron systems at even 
denomenator filling factors}}
\author{S. Sakhi}
\centerline{{\sl Department of Physics and Astronomy, University of
British Columbia,}}
\centerline{{\sl 6224 Agriculture Road, Vancouver, B.C. V6T 1Z1, Canada}}

\abstract{
An interacting double layer system, with uniform positive background, is 
studied at finite temperature in the presense of a strong magnetic field 
corresponding to half filling in each layer. By mapping this system to 
composite fermions in zero field, we investigate the momentum transfer 
rate between the layers. As a result of the residual gauge 
interactions, it turns out that the low temperature 
dependence of this rate is $T^{4/3}$ which is enhanced as compared to the 
normal $T^2$ behavior.}

\endpage

\REF\PraGir{R.~Prange and S.~Girvin, {\sl The quantum Hall effect}
(Springer -Verlag), New York, 1987.
}

\REF\MacPlaBoe{A.M.~MacDonald, P.M.~Platzman, and G.S.~Boebinger, \PRL 
65, 775, 1990; L.~Brey, \PRL 65, 903, 1990. 
}

\REF\SuenEis{Y.~W.~Suen, L.W.~Engel, M.~B.~Santos, M.~Shayegan, and
D.C.~Tsui, \PRL 68, 1379, 1992;
J.P.~Eisenstein, G.S.~Boebinger,
L.N.~Pfeiffer, K.W.~West, and Song He, \PRL 68, 1383, 1992. 
}

\REF\tso{H. C. Tso et al, \PRL 68, 2516, 1992;
K. Flensberg and B. Y. Hu, \PRL 73, 3752, 1994;
Yu. M. Sirenko and P. Vasilopoulos, \PRB 46, 1611, 1992;
L. Zheng and A.M.~MacDonald, \PRB 48, 8203, 1993;
E. Shimshoni and S. L. Sondhi, \PRB 49, 11484, 1994;
A. Kamenev and Y. Oreg, \PRB 52, 7516, 1995.
}

\REF\Duan{J.-Mu Duan, \Euphys 29, 489, 1995.
}

\REF\Rojo{A. G. Rojo and G. D. Mahan, \PRL 68, 2074, 1992.
} 

\REF\Sakhi{S. Sakhi and P. Vasilopoulos, submitted to {\sl Phys. Rev. B}.
}

\REF\LopFra{A. Lopez and E. Fradkin, \PRB 44, 5246, 1991;
ibid, \PRB 51, 4347, 1995. }
 
\REF\Halp{B. Halperin, P. A. Lee and N. Read, \PRB 47, 7312, 1993. }

\REF\Wil{B. L. Altshuler and L.B. Ioffe, \PRL 69, 2979, 1992; 
C. Nayak and F. Wilczek, \NPB 417, 359, 1994;
Y. B Kim, A. Furusaki, X.-G Wen and P. A. Lee, \PRB 50, 17917, 1994. } 

\REF\Ste{Y. B Kim, P.A. Lee, X.-G Wen and P.C.E. Stamp, \PRB 51, 
10779, 1995; A. Stern and B. Halperin, \PRB 52, 5890, 1995. }

\REF\Kim{Y. B Kim and A.J. Millis, Cond-mat/9611125.
}

Two dimensional electron gases involving Coulomb interaction and strong
magnetic field are associated with the observation of many exciting
phenomena such as the integer and fractional quantum Hall effect [\PraGir]. 
Recently
more interest has focused on double-layer two-dimensional electron systems
which exhibit among many others [\MacPlaBoe,\SuenEis], the Coulomb drag 
[\tso]. This phenomenon
manifests when a current flowing in one layer induces a current or voltage
in another nearby layer. It is measured by the transresistivity
$\rho^t=E_{\rm ind}/J_d$ which can be related to the drag rate $1/\tau_D=n
e^2\rho^t/m$ of momentum transfer from one layer to the other. This effect
originates from friction associated with the density fluctuations and
varies as $T^2$ in zero magnetic field. When disorder is present, the drag 
rate is enhanced and
the low temperature dependence changes to $T^2\ln T$. In this work we
investigate further deviations in the drag rate when a strong magnetic
field is present. Specifically, we consider magnetic fields corresponding
to even denomenators filling of Landau levels. 
This situation
is special in the sense that the problem can be mapped to the one of
composite fermions in zero magnetic field at the expense of introducing
residual gauge interactions between them. In such a procedure, though, the
phase space available is drastically modified
giving rise to deviations in the drag rate from the more familiar for
electrons at zero magnetic field. A different drag rate in the presence 
of a magnetic field has been investigated in [\Duan] using flux 
attachment in a bosonic picture. This effect is linear in the 
inter-layer coulomb potential and does not originate from dissipation 
since it is derived from the real term in the total energy.  
Similar effect was found in [\Rojo] to be due to Van 
der Waals forces between the planes. For integer filling and in the 
diffusive regime  we found in [\Sakhi] a drag rate linear in the interlayer  
interaction. Corrections beyond RPA to the drag 
rate were also analyzed in [\Sakhi], these were shown to be important for 
low electronic densities and small layer separation.

Composite fermions are realized through a transformation of the initial
electron system by attaching an even number of flux quanta to each
particle. Mathematically, this is achieved by coupling the electrons to a
statistical gauge field whose dynamics is governed by a Chern-Simons
kinetic term. The resulting fermions experience, in the mean field, an
effective magnetic field which is the sum of the external one and the
statistical magnetic part. This formulation has been analyzed in
[\LopFra] where the response functions and collective modes were
obtained. For $\nu=1/2$ the composite fermions experience a
zero effective magnetic field, and appear to form a fermi surface with
coupling to the fluctuations in the gauge field.  This latter situation
has been analyzed in [\Halp] within a RPA scheme which shows
that the gauge interactions give rise to singular contribution in the
single particle self-energy exhibiting non-Fermi liquid behavior. 
Several other studies were conducted on this compressible state 
for the most part motivated by the large enhancement in the 
effective mass caused by the transverse gauge fluctuations [\Wil,\Ste].

In this work we investigate the effect caused by the above mentioned 
gauge fluctuations on the momentum transfer rate experienced by the 
fermions. We consider a system of two layers of two-dimensional electrons in
the presence of a strong magnetic field $B_{\rm ext}$. We assume the two
layers to be far away from each other that direct transfer of charge 
between them is not possible but include explicitly the coulomb 
interaction between carriers from different layers.

The action at finite temperature of the system is
$$\eqalign{
S&=\int_0^{1/T}d\tau \int d\bfx\sum_{\alpha}\left\{
\psi_{\alpha}^\dagger \left(\partial_\tau +ia_\tau^\alpha-\mu_{\alpha} 
\right)\psi_{\alpha} -{1\over 2m^*}|{\bf D}\psi_{\alpha}|^2\right\}\cr
&-{i\over 2}\sum_{\alpha\beta}\int_0^{1/T}d\tau \int d\bfx 
{\eta}_{\alpha\beta}\epsilon_{\mu\nu\lambda}a_\mu^\alpha\partial_\nu 
a_\lambda^\beta \cr
 &+{1\over 2}\int_0^{1/T} \int d\bfx d\bfy\,\sum_{\alpha\beta}
\,\left(|\psi_{\alpha}({\bf x},\tau)|^2-\bar{\rho_{\alpha}}\right)
U_{\alpha\beta}(|{\bf x-y}|)
\left(|\psi_{\beta}({\bf y},\tau)|^2-\bar{\rho_{\beta}}\right)\cr}
\eqno(1)$$
where the indices $\alpha=1,2$ and $\beta=1,2$ label the layers,
$\bar{\rho_{\alpha}}$ is the average particle density in layer $\alpha$,
and $\mu_\alpha$ is the chemical potential.  ${\bf D}$ 
is the  the covariant
derivative which couples the fermions to the external electromagnetic field
${\bf A}_{\rm ext}$ and to the Chern-Simons field ${\bf a}^\alpha$. The last
term
describes the intra- and inter-layer Coulomb repulsion between the charged
fermions,
$$
U_{\alpha\beta}(|{\bf r}|)=e^2/\epsilon_0 \sqrt{{\bf
r}^2+d^2(1-\delta_{\alpha\beta})}
\eqno(2)$$
and $d$ is the interlayer separation.

The Chern-Simons coefficient in matrix form [\LopFra] is given for 
identical layers by 
$$
{\eta}_{\alpha\beta}={1\over 2\pi(4s^2-n^2)} 
\left(\matrix{2s& -n\cr
               -n&2s\cr}\right);
\eqno(3)$$
where $n, s$ are integers.

Note that by varying the action with respect to $a_\tau^\alpha$ we obtain 
a constraint equation relating the charge density in each layer to the 
statistical magnetic field,
$$
e|\psi_\alpha|^2=\sum_{\beta}{\eta}_{\alpha\beta}b_{\beta}
\eqno(4)$$
It states that fermions in one layer experience a 
statistical flux per particle of $4\pi s$ for the particles in their own 
plane and another flux of $2\pi n$ for the particles in the other layer.
This constraint can be used to render the Coulomb interaction term in (1) 
completely independent of the fermion fields with the advantage of 
allowing a complete integration over them. 

In the mean field approximation, the electrons in each layer experience 
an effective field given by
$$\eqalign{
B_{\rm eff}^\alpha&=B_{\rm ext}+b_\alpha \cr
&=B_{\rm ext}+e\sum_{\beta}{\eta}_{\alpha\beta}^{-1}\bar{\rho_{\beta}} 
\cr} \eqno(5)$$
When $B_{\rm eff}^\alpha$ is such that an integer number $p$ of effective 
landau levels is filled, the filling fraction in each layer 
$\nu_1=\nu_2=p/(1+2lp)$ with $2l=2s+n$. 

In this work we are interested in the situation when the effective field
in each layer vanishes
$B_{\rm eff}^\alpha=0$ $(p\to \infty)$, this corresponds to composite 
fermions forming a
fermi surface with residual gauge interactions $\delta a_\mu^{\alpha}$
describing fluctuations around the mean field state.

Our goal is to obtain the drag rate experienced by the composite fermions 
in each layer and to compare it to the more familiar one for electrons.
In order to find this rate, we probe the system with two gauge 
fields denoted by ${\tilde A}_\mu^\alpha$, and find the 
inter-layer current-current correlation function.
The latter is obtained from the effective action of the system after 
integrating over the fermions and the fluctuating gauge fields.
$$\eqalign{
\exp(-W[{\tilde A}_\mu^\alpha])&=\int {\cal D}\psi {\cal D}\psi^\dagger 
{\cal D}\delta a_\mu^\alpha e^{S}  \cr
&=\int{\cal D}\delta a_\mu^\alpha e^{{\rm Tr}\ln{\bigl(\partial_\tau+{1\over 
2m^*}\bigl(-i\nabla-e{\tilde A^\alpha}-\delta 
a^\alpha\bigr)^2-\mu_\alpha+i\delta 
a_\tau^\alpha\bigr)}-{1\over 2}\int \delta b^\alpha 
{\tilde U}_{\alpha\beta}\delta b^{\beta}-i\int \delta 
a_\tau^\alpha {\eta}_{\alpha\beta}\delta b^{\beta}} \cr}
\eqno(6)$$
where  ${\tilde U}_{\alpha\beta}(q)=[{\eta}^tV{\eta}]_{\alpha\beta}$ and 
the ${\rm Tr}
\ln(...)$ results from fermionic integrations. We find it more convenient to
work in the coulomb gauge for both
the  Chern-Simons gauge fields and the probing fields. This allows the 
following parametrization $\delta a_i(q)=i\epsilon^{ij}q_j \phi(q)$ and 
${\tilde A_i}(q)=i\epsilon^{ij}q_j \phi_0(q)$.

The strategy is to expand the ${\rm Tr}\ln(...)$ in (6) 
up to 
quadratic fluctuations in $\delta a_\tau^\alpha$ and $\phi^\alpha$, and then 
carry out an integration over these degrees. The outcome of such procedure
is the effective action describing the probing electromagnetic fields. The
current-current 
correlation function relevant for the drag rate is achieved by double 
functional differentiation with respect to $\phi_0^1$ and $\phi_0^2$. 

At an intermediate stage we obtain the following action,
$$\eqalign{
S_{\rm eff}&={1\over 2}\int (\phi^\alpha(-q)\qquad \delta a_\tau^\alpha(-q))
\left(\matrix{\Pi_v\delta_{\alpha\beta}+W_{\alpha\beta}& 
iq^2{\eta}_{\alpha\beta}\cr  
iq^2{\eta}_{\alpha\beta}&\Pi_s\delta_{\alpha\beta}\cr}\right) 
\left(\matrix{\phi^{\beta}(q)\cr \delta a_\tau^{\beta}(q)  \cr}\right) \cr
&+i\int_p\int_q \phi_0^\alpha(-p-q)\phi^\alpha(q)\delta 
a_\tau^\alpha(p)\Delta(p,q)\cr
&+i\int_p\int_q \phi_0^\alpha(-p-q)\delta a_\tau^\alpha(p)\delta 
a_\tau^\alpha(q)\Lambda(p,q)\cr
&+i\int_p\int_q 
\phi_0^\alpha(-p-q)\phi^\alpha(q)\phi^\alpha(p)\Gamma(p,q) \cr} \eqno(7)$$
here $\int_q$ is a shorthand notation for $T\sum_\omega\int
d\bfq/(2\pi)^2$ and $W_{\alpha\beta}=q^4{\tilde U}_{\alpha\beta}$. The 
polarization 
functions are given in the long wavelength and low frequency as 
$\Pi_s(q,\omega)=m^*/2\pi$, $\Pi_v(q,\omega)=q^4({1\over 12\pi m^*}+
\gamma {|\omega|\over q^3})$ and $\gamma=2n/m^*v_F$ [\Halp]. Note 
that there is no induced Chern-Simons term since the composite fermions do not experience any 
magnetic field. The vertex functions are 
defined in terms of the single-particle Green functions $G$ as 
$$
\eqalignno{
\Lambda(p,q)&={1\over 4m^*}\int_k 
k\wedge(p+q)G(k)G(k+p+q)\left[G(k+p)+G(k+q)\right] &(8a) \cr
\Delta(p,q)&={1\over m^{*2}}\int_k 
k\wedge(p+q)G(k)G(k+p+q)\left[((k+p)\wedge q) G(k+p)+(k\wedge q)G(k+q)\right]
& (8b) \cr
\Gamma(p,q)&={1\over 2m^{*3}}\int_k (k\wedge(p+q))G(k)G(k+p+q)\cr
&\times\left[
(k\wedge p) ((k+p)\wedge q)G(k+p)+ (k\wedge q)((k+q)\wedge p)G(k+q)\right]
&(8c)\cr } $$
here $k\wedge p=\epsilon^{ij}k_ip_j$. The gauge propagators are obtained 
by inverting the matrix form in (7). We
give here those coupling the two layers which are  relevant  to drag rate.
They are given by 
 $$\eqalignno{
{\cal D}_{\phi\phi}^{12}(q)&= {2\pi e^2 {\eta}^2e^{-qd}\over q\left( 
{\gamma|\omega|\over 
q}+\chi_sq\right)\left( {\gamma|\omega|\over q}+\chi_dq^2\right)} &(9a)\cr
 {\cal D}_{\tau\phi}^{12}(q)&=\left({2\pi {\eta}^3\over m}\right){2\pi e^2 
q^2e^{-qd}\over \left( {\gamma|\omega|\over
q}+\chi_sq\right)\left( {\gamma|\omega|\over q}+\chi_dq^2\right)}  & (9b)\cr
{\cal D}_{\tau\tau}^{12}(q)&=\left({2\pi {\eta}^2\over m}\right)^2{2\pi e^2 
q^3e^{-qd}\over \left( {\gamma|\omega|\over 
q}+\chi_sq\right)\left( {\gamma|\omega|\over q}+\chi_dq^2\right)}   
&(9c)\cr } $$
with $\chi_s=4\pi {\eta}^2e^2$, $\chi_d={1\over 12\pi m^*}+{2\pi 
{\eta}^2\over
m^*}(1+m^*e^2d)$. In deriving the gauge propagators we considered the 
special case of no inter-layer flux attachment ${\eta}_{12}=0$, this is done 
since we are more concerned with the effect 
arising from the inter-layer coulomb interaction. The inclusion of such 
a term would add an effective attraction between the two layers. While 
${\eta}={1\over 4\pi}$ for $\nu=1/2$, it is useful to keep it
as a variable describing the fermion-gauge coupling. The limit ${\eta}\to
\infty$ should reproduce the already known results in the absence of
statistical gauge interaction.

Eliminating the gauge fluctuations by a gaussian integration results in an
effective action for the probing electromagnetic field 
$W[\phi_0^\alpha]$. The transconductivity is obtained from the cross quadratic
term, in Matsubara frequencies it is found to be
$$\eqalign{
\sigma^{t}(\bfq,\omega_q)&={2\over q^2\omega_q} {\delta^2W\over \delta 
\phi_0^1(-q)\delta \phi_0^2(q)}\cr &={-2\over
q^2\omega_q} \int_p {\cal D}_{\tau\tau}^{12}(p){\cal D}_{\tau\tau}^{12}(p+q)\Lambda(-p,p+q)\Lambda(p,-p-q)\cr
&-{1\over q^2\omega_q}\int_p  \left[{\cal D}_{\tau\tau}^{12}(p){\cal 
D}_{\phi\phi}^{12}(p+q)+ {\cal 
D}_{\tau\phi}^{12}(p) {\cal D}_{\phi\tau}^{12}(p+q)\right]\Delta(-p,p+q)\Delta(p,-p-q) \cr
&-{2\over
q^2\omega_q} \int_p {\cal D}_{\phi\phi}^{12}(p){\cal D}_{\phi\phi}^{12}(p+q) \Gamma(-p,p+q)\Gamma(p,-p-q) \cr}
\eqno(10)
$$
the first contribution is the only one that survives in the limit  ${\eta}\to
\infty$, it describes the usual drag rate due to density fluctuations,
this has been analyzed by several  authors. The two others  involve the 
singular transverse gauge field. Since the most important  
frequencies involved in (9a) are of order $\omega\sim q^3$, the last 
contribution in (10) is the 
most likely to give marked deviation in the temperature dependence of 
the drag rate. For this reason we give more details on this contribution, 
similar manipulations can be carried on the other two. First, we take 
$q=0$ and 
using a contour integration followed by an analytical continuation to 
real frequencies $i\omega_q\to \omega+i0$, the transconductivity can be
written as
$$\eqalign{
\sigma^t(\omega+i0)&={-2\over \pi\omega}\int {d^2p\over
(2\pi)^2}\int_{-\infty}^{+\infty}d\epsilon\left[f(\epsilon+\omega)-
f(\epsilon)\right] {\cal D}_{\phi\phi}(p,\epsilon){\cal
D}_{\phi\phi}(p,\epsilon+\omega){\tilde 
\Gamma}^{-+}(p;\epsilon,\epsilon+\omega)\cr &\times
{\tilde \Gamma}^{+-}(p;\epsilon+\omega,\epsilon)\cr &+f(\epsilon){\cal 
D}_{\phi\phi}(p,\epsilon){\cal D}_{\phi\phi}(p,\epsilon+\omega)
{\tilde \Gamma}^{++}(p;\epsilon,\epsilon+\omega) 
{\tilde \Gamma}^{++}(p;\epsilon+\omega,\epsilon)\cr &-
f(\epsilon+\omega)){\cal D}_{\phi\phi}(p,\epsilon){\cal
D}_{\phi\phi}(p,\epsilon+\omega){\tilde  
\Gamma}^{--}(p;\epsilon,\epsilon+\omega)
{\tilde \Gamma}^{--}(p;\epsilon+\omega,\epsilon)\cr} \eqno(11)$$
where $f(\epsilon)=(\exp(\epsilon/T)-1)^{-1}$, the $+-$ indicate the part in 
the complex plane where the vertex functions are analytically continued.

The vertex function ${\tilde \Gamma}=\lim_{q\to 0}\Gamma/q$ is derived from 
(8c) using the assumption of a finite scattering rate $1/\tau$ in the 
single-particle Green functions. For zero external frequency we find 
$\Gamma^{+-}(p;\epsilon,\epsilon)\sim v_F\tau\epsilon p_1 p/4\pi$, and 
$\Gamma^{++}(p;\epsilon,\epsilon)=0$. The {\it d.c} 
transresistivity defined as 
$\rho^t=\sigma^t/(\sigma^2+\sigma_t^2)\approx\sigma^t/\sigma^2$, where 
the in-plane conductivity $\sigma=ne^2\tau/m$, is then
$$ \rho^t={1\over 32\pi^4e^4\gamma^2}\int_{-\infty}^{\infty} 
{d\epsilon\over T}{\epsilon^2\over \sh^2\left({\epsilon\over 
2T}\right)}\int dp p^5 \left[{\cal D}_{\phi\phi}(p,\epsilon)\right]^2
\eqno(12)
$$

The momentum integration is cut off in the ultra-violet by $1/d$ and the 
integrand is important for $\epsilon\sim T_0(pd)^3$ with the temperature 
scale $T_0=\chi_d/\gamma d^3$. A simple evaluation in that region with 
the replacement $\epsilon^2/ \sh^2(\epsilon/2T)$ by $4T^2$ for 
$|\epsilon|\leq T$ leads to 
$$
\rho^t= {1\over (e\gamma d)^4}(T/T_0)^{4/3}F\left((T_0/T)^{2/3}\right) 
\eqno(13)$$
where the scaling function is $F(t)={1\over 192\pi^3\sqrt{3}}[1+{6\over 
\pi}\tan^{-1}[(2t-1)/\sqrt{3}]-{\sqrt{3}\over 
\pi}\ln[(t+1)^2/(t^2-t+1)]]$. 

This results in a low-temperature dependence in the drag rate 
$1/\tau_D\sim T^{4/3}$ which is enhanced as compared to the usual 
drag rate [\tso]. In fact, the latter involving the vertex function (8a) 
can be obtained using the same manipulations. In this case, one finds 
${\tilde 
\Lambda}^{+-}(p;\epsilon,\epsilon)=(\tau/2m^*)\Im\Pi(p,\epsilon+i0)\approx 
-\epsilon\tau/(8\pi pv_F)$ and $\rho^t=[\eta^4/(48\pi p_F^4e^4\gamma^6 
d^6)](T/T_0)^2\ln(T_0/T)$. The limit $\eta\to \infty$, representing 
gauge-fermion decoupling leads to $\rho^t=(T^2/p_F^4e^8\gamma^4 
d^2)\ln(T_0/T)$.   

We have shown that transverse gauge fluctuations in a double-layer system 
lead to deviations in the temperature dependence of the drag rate of 
momentum transfer. It is a direct consequence of the 
relaxation rate for the long-wavelength difference charge density between the 
layers that varies as $q^3$.

\ACK

I acknowledge useful discussions with P. Vasilopoulos which lead to this 
analysis. This work was supported by the Natural Science and  
Engineering Research Council of Canada.

\noindent   
{\it Note added}, it was brought to my attention that a similar 
$T^{4/3}$ dependence was found in [\Kim] where a modified Drude formula 
was adopted.
\refout

\bye